%% file: ms.tex
\documentclass[conference]{IEEEtran}
\IEEEoverridecommandlockouts
\usepackage{amsmath,amssymb,amsfonts}
\usepackage{algorithmic}
\usepackage{graphicx}
\usepackage{textcomp}
\usepackage{xcolor}
\usepackage{xspace}
\usepackage{subcaption}
\usepackage[hidelinks]{hyperref}
\usepackage{soul}
\usepackage{relsize}

\usepackage{todonotes}

\def\BibTeX{{\rm B\kern-.05em{\sc i\kern-.025em b}\kern-.08em
    T\kern-.1667em\lower.7ex\hbox{E}\kern-.125emX}}

\newcommand{\frsz}{FRSZ2\xspace}
\newcommand{\frszaccN}[1]{frsz2\_#1}
\newcommand{\frszacc}[1]{\frszaccN{#1}\xspace}
\newcommand{\doubleN}{float64}
\newcommand{\floatN}{float32}
\newcommand{\halfN}{float16}
\newcommand{\double}{\doubleN\xspace}
\newcommand{\float}{\floatN\xspace}
\newcommand{\half}{\halfN\xspace}
\newcommand{\Acc}[1]{Acc\textless#1\textgreater\xspace}

\begin{document}

\title{\frsz for In-Register Block Compression Inside GMRES on GPUs
\thanks{
We thank the U.S. Department of Energy and the National Science Foundation for funding this work.
}
}

\author{
\IEEEauthorblockN{Thomas Gr\"utzmacher\IEEEauthorrefmark{1} \and Robert Underwood\IEEEauthorrefmark{2} \and Sheng Di\IEEEauthorrefmark{2} \and Franck Cappello\IEEEauthorrefmark{2} \and Hartwig Anzt \IEEEauthorrefmark{3}}\and%
\IEEEauthorblockA{\IEEEauthorrefmark{1}\textit{Karlsruhe Institute of Technology} \\
\textit{Technical University of Munich} \\
thomas.gruetzmacher@tum.de}\and%
\IEEEauthorblockA{\IEEEauthorrefmark{2}\textit{Argonne National Laboratory} \\
\textit{University of Chicago}\\
\{runderwood, sdi1, cappello\}@anl.gov}\and%
\IEEEauthorblockA{\IEEEauthorrefmark{3}\textit{Technical University of Munich} \\
\it{University of Tennessee, Knoxville}\\
hartwig.anzt@tum.de}%
}

\maketitle

\begin{abstract}
The performance of the GMRES iterative solver on GPUs is limited by the GPU main memory bandwidth. Compressed Basis GMRES outperforms GMRES by storing the Krylov basis in low precision, thereby reducing the memory access. An open question is whether compression techniques that are more sophisticated than casting to low precision can enable large runtime savings while preserving the accuracy of the final results. This paper presents the lightweight in-register compressor \frsz that can decompress at the bandwidth speed of a modern NVIDIA H100 GPU. In an experimental evaluation, we demonstrate using \frsz instead of low precision for compression of the Krylov basis can bring larger runtime benefits without impacting final accuracy.
\end{abstract}



\input{sections/introduction} 
\input{sections/background} 
\input{sections/problem_formulation} 
\input{sections/design} 
\input{sections/methodology} 
\input{sections/evaluation} 
\input{sections/related_work} 
\input{sections/conclusion} 

\section*{Acknowledgment}

This work was performed on the NHR@KIT Future Technologies Partition testbed funded by the Ministry of Science, Research and the Arts Baden-Württemberg and by the Federal Ministry of Education and Research.
The authors thank the Innovative Computing Lab for access to their NVIDIA H100 GPU.

This work was supported by the U.S. Department of Energy, Office of Science, Advanced Scientific Computing Research Program, under Contract DE-AC02-06CH11357.
This research was supported by the Exascale Computing Project (17-SC-20-SC), a collaborative effort of the U.S. Department of Energy Office of Science and the National Nuclear Security Administration.
This material is based upon work supported by the National Science Foundation under Grant No. \#2311875 and \#2104203.
This work has been conducted within the Joint Laboratory for Extreme-Scale Computing (JLESC).


\input{ms.bbl}
\end{document}

%% file: sections/introduction.tex
\section{Introduction}
The Generalized Minimal Residual Method (GMRES) is a popular method for solving linear systems of equations iteratively. GMRES is widely used in applications that result in large, sparse linear systems that are not symmetric positive definite. Systems of this kind are common in scientific computing applications, ranging from finite element discretizations over combinatorial problems to circuit simulations. GMRES builds up out of matrix-vector operations with the system matrix, vector operations, and orthogonalization, building up the Krylov subspace and a minimization process. All these building blocks are memory-bound, and thus the performance of the GMRES solver is limited by the main memory bandwidth of the processor.
Consequently, strategies to accelerate the GMRES method aim to reduce the data transfer. In the Compressed Basis GMRES (CB-GMRES~\cite{cbgmres}) method, the Krylov basis vectors are compressed by conversion to lower precision formats, e.g., IEEE 754 single precision or half precision. This strategy reduces the data transfers of the individual iterations and allows for faster execution of the iterations while mostly preserving the quality of the final solution. The information loss caused by storing the Krylov basis in low precision can delay convergence, experiments however indicate that in most scenarios the convergence delay is easily compensated by the faster exection~\cite{cbgmres}.
Converting the individual vector values to low precision is a straightforward compression scheme, and a valid question is whether more sophisticated compression schemes operating not on a value level but on a block level allow for higher compression rates or reduced information loss. 

In this paper, we investigate this question by employing lossy compression for the compression of the Krylov vectors inside the CB-GMRES algorithm.
Lossy compression maps the input data into a different representation with a smaller memory footprint through a series of computations that decorrelate and then encode common patterns in the data leaving a smaller representation.
However, the compression has to happen in processor registers and needs to be extremely fast to not incur measurable overhead to the Krylov solver. 
In particular, all compression and decompression have to be hidden behind the memory access. A quick pen-and-paper calculation reveals this strategy to be viable: The latest Nvidia server line GPU, the Nvidia H100 GPU, has a memory bandwidth of roughly 2 TB/s and a peak double-precision performance of $\approx 25$ TFLOP/s.
That means an algorithm can execute up to 100 double-precision (64-bit) computations per double-precision value retrieved from main memory before hitting the compute peak.
For a sparse linear algebra routine executing 4 operations on each retrieved value, one could consider 96 operations to be ''wasted''. Suppose some of these operations could be used to compress the data to consume only 32 bits per value. In that case, the compute-to-read ratio reduces to 50:1 and one could use 46 operations for compression and decompression of the double-precision values. While this sounds viable, several critical aspects constrain the design of the compression strategy.
First, implementing compression and decompression in only 46 operations is not straightforward and excludes many sophisticated compression strategies. 
Second, the compressor must support at least random access by block to support the memory access patterns used within CB-GMRES.
Third, the data values to be compressed in CB-GMRES are generally uncorrelated, presenting a challenge to effectively decorrelate.

This paper presents \frsz, a highly specialized compressor designed to support CB-GMRES.
Our key contributions include:
\begin{enumerate}
    \item We study the impacts of lossy compression error bounds on this problem to demonstrate that this problem prefers point-wise error bounds.
    \item We identify the bottlenecks of using compression techniques for this problem and accommodate these in the compressor design.
    \item We describe in detail the design of \frsz and demonstrate how it integrates into Ginkgo's CB-GMRES.
    \item We compare \frsz to other compression techniques and demonstrate that \frsz is the fastest compressor, achieving performance 99.6\% of loading double precision from memory, which is $1.2\sim3.1\times$ faster than the next fastest compressor cuSZp2 at the roofline.
    \item We report that for selected problems, \frsz compression can render performance advantages of up to 1.3$\times$ over single-precision compression in CB-GMRES.
    \item We discuss how lossy compression could be used in a generic framework for accelerating CB-GMRES.
\end{enumerate}

The remainder of the paper is organized as follows:
In Section~\ref{sec:background}, we describe CB-GMRES and describe in detail where compression can be applied.
In Section~\ref{sec:problem_formulation} and Section~\ref{sec:design}, we present the challenges and how we tackle them in the design of \frsz. 
We then present our evaluation methodology in Section~\ref{sec:methodology} before presenting the experimental results in Section~\ref{sec:evaluation}.
We provide an overview of related work in Section~\ref{sec:related_work} and conclude in Section~\ref{sec:conclusions} with a summary of the findings, challenges, and potential.

%% file: sections/background.tex
\section{Background}
\label{sec:background}
Krylov methods are an essential building block in scientific high-performance computing for the rapid solution of large, sparse linear systems.
They approximate the solution in a subspace that is generated iteratively in the Krylov solver iteration. Starting from an initial guess, each iteration adds a Krylov basis vector by orthogonalizing a new search direction against the already computed basis until the subspace spanned by the basis is large enough to contain a solution approximation of sufficient accuracy.
For a problem of dimension $n$, the exact solution is available after $n$ iterations, as then the Krylov basis spans the whole space.
However, in practice, a much smaller number of iterations is typically sufficient to find a suitable solution approximation.
Long recurrence Krylov methods, like the popular GMRES solver we focus on in this paper, build up the Krylov basis in main memory, and for each new search direction, the pre-existing basis has to be retrieved from main memory for the orthogonalization procedure such that it can then be appended to the extended basis.
This makes the orthogonalization a memory-bound step that often dominates the overall solver runtime.
A strategy to accelerate the Krylov solver can thus be to compress the data of the Krylov basis to reduce the main memory access volume. In~\cite{cbgmres}, the authors realize this idea by storing the Krylov basis in low precision -- a simple lossy compression technique.
The GMRES algorithm and the compression potential is shown in Figure~\ref{fig:gmres}.
Reducing the precision may incur perturbations in the Krylov basis, thereby harming the convergence.
The experimental results, however, reveal that the slower convergence can typically be compensated by faster execution. Hence, more Krylov basis vectors can be generated in less time, thereby still accelerating the solution process.
In this paper, we replace the compression of the vector entries based on casting to lower precision with a more sophisticated block-based compression.
The hope is that higher compression ratios can be achieved while still hiding all compression and decompression behind the memory access. 

\begin{figure}
\begin{center}
\fbox{
\begin{minipage}[t]{0.9\textwidth} 
{\smaller[3]    
\begin{tabbing}
xxx\=xxx\=xxx\=xxx\=xxx\=\kill
\> \itshape  1. \' Compute $r_0 := b - Ax_0$, $\beta := \|r_0\|_2$, and $v := r_0 / \beta$. Set \hl{$V_1=[\,v\,]$}\\
\> \itshape  2. \' \texttt{\bf for} $j := 1, 2, \ldots, m$ \\
\> \itshape  3. \' \> Compute $w := A (M^{-1}v)$ \\
\> \itshape  4. \' \> $\omega := \|w\|_2$ \\
\> \itshape  5. \' \> \hl{Orthogonalize $h_{1:j,j} := V_j^T w$, $w := w - V_j h_{1:j,j}$ } \\
\> \itshape  6. \' \> $h_{j+1,j} := \|w\|_2$ \\
\> \itshape  7. \' \> \texttt{\bf if} $(h_{j+1,j} < \eta \, \omega)$ \texttt{\bf then}\\
\> \itshape  8. \' \> \> \hl{Re-orthogonalize $u := V_j^T w$, $w := w - V_j u$ } \\
\> \itshape  9. \' \> \> $h_{1:j,j} := h_{1:j,j} + u$ \\
\> \itshape 10. \' \> \> $h_{j+1,j} := \|w\|_2$ \\
\> \itshape 11. \' \> \texttt{\bf endif} \\
\> \itshape 12. \' \> \texttt{\bf if} $(h_{j+1,j} = 0)$ \texttt{\bf or} 
                      $(h_{j+1,j} < \eta \, \omega)$ \texttt{\bf then} set $m := j$ and \texttt{\bf go to step 17}, \texttt{\bf endif} \\
\> \itshape 13. \' \> $v := w / h_{j+1,j}$ \\
\> \itshape 14. \' \> Set \hl{$V_{j+1} := \left[V_j,~v\right]$} \\
\> \itshape 15. \' \texttt{\bf endfor} \\
\> \itshape 16. \' Define the $(m + 1) \times m$ Hessenberg matrix $\bar{H}_m = \left(h_{ij}\right)_{1 \le i \le m+1, 1 \le j \le m}$ \\[0.05in]
\> \itshape 17. \' Compute $y_m$ the minimizer of $\|\beta e_1 - \bar{H}_m y\|_2$ and \hl{$x_m := x_0 + M^{-1} (V_m y_m)$} \\
\> \itshape 18. \' \texttt{\bf if} satisfied \texttt{\bf then Stop}, \texttt{\bf else} set $x_0 := x_m$ and \texttt{\bf go to step 1}, \texttt{\bf endif}
\end{tabbing}
}
\end{minipage}
}
\end{center}
\caption{Algorithmic formulation of the restarted GMRES algorithm for solving sparse linear systems. Sections where compression can be used are highlighted.}
\label{fig:gmres}
\vspace{-.5cm}
\end{figure}


%% file: sections/problem_formulation.tex
\section{Problem Formulation}\label{sec:problem_formulation}

Designing block compression inside of GMRES on the GPU presents several critical requirements on the compressor to produce a usable solution: 1) Quality: the compression error must not affect the solution accuracy, 2) Decorrelation: the compression needs to succeed in reducing the data volume without sacrificing too much information, and 3) Performance: The compression and decompression needs to be hidden behind the memory access to not incur any overhead to the CB-GMRES algorithm.

In the following subsections, we expand on the definition of the critical requirements of decorrelation and performance and how they impact the design of a compressor.  We discuss quality later in Section~\ref{sec:evaluation}.

\subsection{Decorrelation}\label{sec:problem_decorrelation}
\begin{figure}
    \centering
    \begin{subfigure}[b]{.5\columnwidth}
        \includegraphics[width=\textwidth]{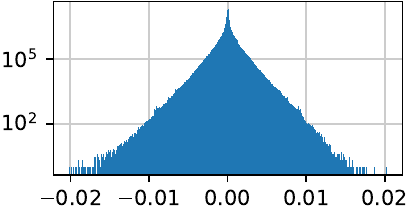}
        \caption{histogram values}\label{fig:dist_values}
    \end{subfigure}%
    \begin{subfigure}[b]{.5\columnwidth}
        \includegraphics[width=\textwidth]{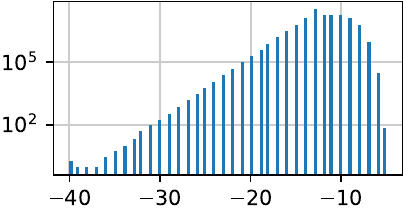}
        \caption{histogram exponent}\label{fig:dist_exp}
    \end{subfigure}\\
    \begin{subfigure}[b]{.5\columnwidth}
        \includegraphics[width=\textwidth]{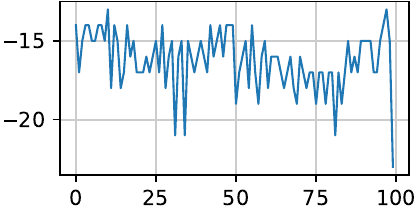}
        \caption{values}\label{fig:values_values}
    \end{subfigure}%
    \begin{subfigure}[b]{.5\columnwidth}
        \includegraphics[width=\textwidth]{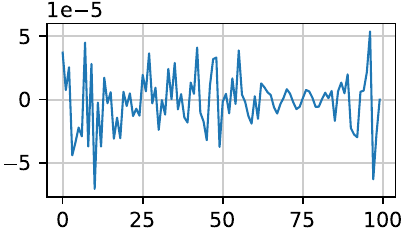}
        \caption{exponent}\label{fig:exp_values}
    \end{subfigure}\\
    \caption{Histogram of Exponents and values from the atmosmodd matrix.  Only the exponent has a few common values, but values are normally distributed making decorrelation difficult.}
\end{figure}

The Krylov vectors compressed in GMRES are difficult to decorrelate.
Krylov vectors are normalized, which means all values are in $[-1,1]$.  Figure~\ref{fig:dist_values}-\ref{fig:exp_values} show the values and distribution of the values.
While the first iteration of a solver may show some patterns in the data to be compressed\footnote{For example, the solver might initialize the vectors using values from a $\sin$ function.}, the values become uncorrelated in the subsequent iterations.
There is no particular pattern to their ordering or values with both uniformly distributed over the domain.

To understand why these vectors are hard to decorrelate and the impact that has on the compressor design, it is helpful to consider how lossy compressors work.
Modern lossy compressors feature three key stages to achieve high compression ratios: decorrelation, quantization, and encoding \cite{cappello_use_2019}.
Decorrelation is a class of techniques that reduces autocorrelation in data and produces a new version with a distribution that ideally has a much smaller variance that can then be quantized into fewer values, finally reducing entropy and allowing improved compression ratios.
Each leading lossy compressor uses different decorrelation mechanisms. For example, SZ uses a collection of predictors (e.g., cubic spline \cite{liang_sz3_2023}, Lorenzo \cite{di_fast_2016}, block linear regression \cite{tao_significantly_2017}) to predict later values with earlier values, and ZFP \cite{lindstrom_fixed-rate_2014} uses a near orthogonal transform similar to JPEG.
However, all of these methods rely on patterns in the values to reduce the entropy.  Figure~\ref{fig:dist_values} reveals that these patterns do not exist for Krylov vectors.
In consequence, the compression will be ineffective at best or counterproductive at worst. In the worst case, an ineffective decorrelation mechanism can introduce systematized decompression errors or increase compressed size \footnote{e.g. from space usage overheads from the unpredictable data correction mechanisms in prediction scheme-based methods like SZ}.

Some compression is still possible, as, for example,  a substantial fraction of the values in the Krylov basis share common exponents, see Figure~\ref{fig:exp_values}.
This inspires a design that attempts to decorrelate the exponents but not the values.
To the best of our knowledge, we present the first design for this kind of decorrelation scheme.

\subsection{Performance}\label{problem_performance}

For this paper, we will define the performance challenge in terms of the speedup relative to the standard GMRES using IEEE 754 double precision for all arithmetic operations and for storing the Krylov basis vectors. Additionally, we will compare against the CB-GMRES algorithm storing the Krylov basis in single(\float)- and half(\half)-precision, respectively.

The performance envelope for compression is more aggressive than converting to single- or half-precision without increased information loss, is extremely tight.
As explained in the introduction, there is time for only about 46 instructions to perform compression and decompression without affecting the runtime of the solver.
This eliminates entire classes of encoding stages that use methods such as Huffman encoding and the embedded encoding used by ZFP that require far too many instructions, leaving only designs with more primitive truncation-based encoding schemes.

Notwithstanding the challenges with the decorrelation schemes used in other ultra-fast GPU compressors discussed previously in Section~\ref{sec:problem_decorrelation}, even the fastest compression schemes with simple encoding schemes such as cuSZp2 \cite{yafan_huang_cuszp2_2024} are far too computationally complex and exceed the instruction limits. Even at its fastest configuration, cuSZp2 achieves only 1241GB/s on an A100 GPU, which is $\approx80\%$ of its bandwidth. In a more typical case, it achieves closer to 500GB/s, which is $\approx32\%$ of its bandwidth, making it too slow to be used in GMRES without an unacceptable slowdown. 

Consequently, a highly specialized high throughput compression algorithm is required to compete with compression based on converting to low precision.

%% file: sections/design.tex



\section{Design}
\label{sec:design}


In this section, we detail the new compression format \frsz that is fast in decompression while retaining more information per value than IEEE \float for GMRES.
In order to decorrelate the exponent information of values efficiently, the format initially evaluates the universally used IEEE 754 double-precision format \double.
Each \double can be separated into its sign $s$, 11-bit unsigned exponent $e$, and $52$ significand bits $b_{51}\dots b_1b_0$.
Formula~\ref{eq:ieee_value} is used to compute the represented value for the most common format:
\begin{equation}
\label{eq:ieee_value}
\text{value} \, = \, (-1)^s\cdot (1.b_{51}\dots b_1b_0)_2 \cdot 2^{e-1023}
\end{equation}

The exponent is stored in an offset-binary representation, which means it is stored as an unsigned integer and needs to be subtracted by an offset number to get the actual value.
For \double, this offset is $1023$.
The significand in (Equ. \ref{eq:ieee_value}) has a leading $1$ bit, which is not explicitly stored.

The compression aims to group multiple values into a block and extract their exponent \footnote{block floating point implementations are not novel.  They are used in ZFP \cite{lindstrom_fixed-rate_2014} and proposed as early as 1964 in \cite{wilkinson_rounding_1964}.  We differ from ZFP in that we do not feature a decorrelation stage which is counterproductive for this data and our unique data layout.  Our approach differs from \cite{wilkinson_rounding_1964} in our data layout}.
Freely choosing which values to group is impossible because that would reduce our decompression speed for consecutive values.
The idea is that neighboring Krylov vector values are likely close in magnitude, which means they can be grouped together.
The block size is fixed to avoid global synchronization points and increase the throughput on GPUs, which are massively parallel architectures.
This block size BS is one of two optimization parameters of \frsz.
An effective value will be determined in Section~\ref{sub:implementation}.

To account for small differences in the magnitude of values in a block, the maximum IEEE exponent $e_\text{max}$ is identified, and all values are normalized with this exponent.
This implies that, in contrast to the IEEE format, the significands are not normalized to sub-unit values, but the integer part of the significand of the represented numbers has to be stored.
For each number in a block that has a smaller exponent $e < e_\text{max}$, we need to prefix the significand with $k = e_\text{max} - e$ zeros in order to represent the same value.
We limit the number of bits per value, which includes the sign and significand, to a fixed length $l$.
$l$ adjusts the compression ratio and the maximum precision retained per value.
We evaluate and advocate for different $l$ in Section~\ref{sub:implementation}.

We represent the compressed value, consisting of the sign bit and the significand, with the symbol $c$.
For $l$ bits, the compressed value $c_{l-1}\dots c_1c_0$ represents the following number:
\begin{equation}
\label{eq:frsz2_value}
\text{value} \, = \, (-1)^{c_{l-1}}\cdot (c_{l-2}.c_{l-3}\dots c_1c_0)_2 \cdot 2^{e_\text{max}}
\end{equation}

To summarize, the sign bit is stored as the most significant bit of $c$, followed by the significand's integer part and then the significand's fractional part.
BS and $l$ are the two optimization parameters of \frsz.
For increased memory access speed, we read and write our memory as integers with at least $l$ bits, which requires the beginning of a block to be aligned to that value.
The exponent is stored in integer representation.
For $16 < l \leq 32$, we use an integer type with 32 bits, so the memory needs to be aligned to 4 bytes.
The storage requirement in bytes for $n$ elements and an assumed integer representation with 4 bytes is:
\begin{equation}
\label{eq:frsz_storage}
\underbrace{
    \left \lceil{\frac{n}{\text{BS}}} \right \rceil  \cdot \left \lceil{ \frac{\text{BS} \cdot l}{4} }\right \rceil \cdot 4
}_{\text{compressed values}}
+ \underbrace{
      \vphantom{\left \lceil{\frac{n}{\text{BS}}} \right \rceil  \cdot \left \lceil{ \frac{\text{BS} \cdot l}{4} }\right \rceil \cdot 4} 
      \left \lceil{\frac{n}{\text{BS}}} \right \rceil \cdot 4
  }_{\text{exponents}}
\end{equation}



\subsection{Compression}

\begin{figure}[htb]
    \centering
    \includegraphics[width=\columnwidth]{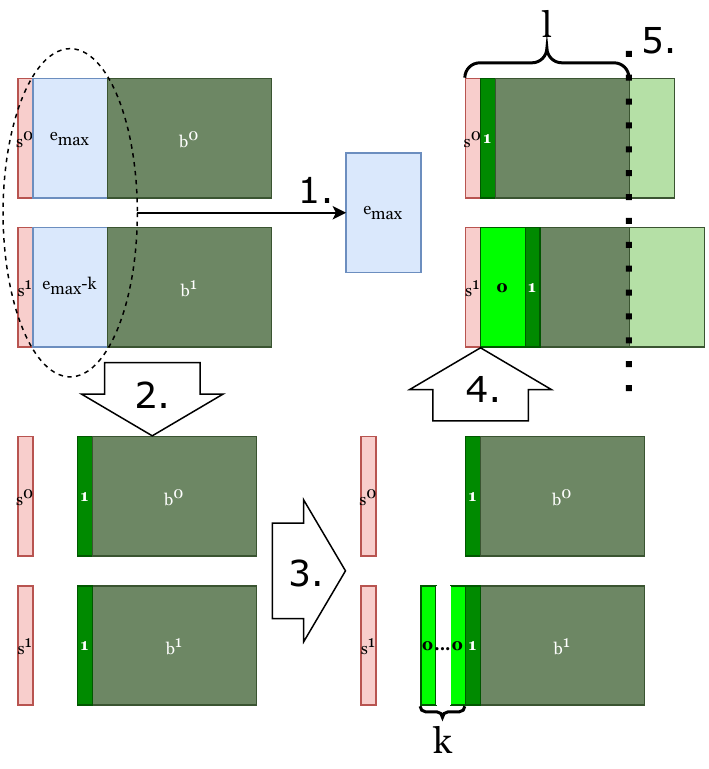}
    \caption{\frsz compression steps (BS $= 2$ and arbitrary $l>2$).}
    \label{fig:frsz_visualization}
\end{figure}

The compression algorithm performs the following~\ref{comp:store} steps: 
\begin{enumerate}
    \item Extract the exponent $e$ and find the maximum exponent $e_\text{max}$ from all values in the block;
        \label{comp:maximum}
    \item Extract the sign $s$ and significand; add the usually implicit $1$ bit to the significand representation;
        \label{comp:significand}
    \item Normalize the significand to the maximum exponent $e_\text{max}$ by prefixing the significand with $k = e_\text{max}-e$ $0$ bits;
        \label{comp:normalize}
    \item Put the sign bit to the left of the normalized significand;
        \label{comp:merge}
    \item Cut the new representation to the appropriate length $l$. Now, we have $c$;
        \label{comp:cut}
    \item Store $e_\text{max}$ and all $c$ of the block.
        \label{comp:store}
\end{enumerate}
An illustration of this process is provided in Figure~\ref{fig:frsz_visualization}.
If $l$ does not match the size of the integer representation type exactly (e.g. $l = 21$), step~\ref{comp:store} needs to merge neighboring values before storing them since GPUs can only store values at a byte level.

The compression must be performed on all BS elements simultaneously to efficiently utilize the GPU bandwidth.
Updating just a single element would require reading $e_\text{max}$ before writing the compressed value.
If $e_\text{max}$ changes as a result, all values of the same block need to be read from memory, renormalized to this new exponent, and written back to memory.

\subsection{Decompression}

Decompression is an easier procedure and does not require the full block to be read simultaneously.
The following steps retrieve a value at index $i$:
\begin{enumerate}
    \item Read $e_\text{max}$ for the corresponding block and read the correct compressed value $c$ at index $i$;
        \label{decomp:read}
    \item Separate $c$ into the sign bit $s$ and the significand; Count the number of inserted zeros $k$ at the beginning of the significand;
        \label{decomp:count_zero}
    \item Remove the inserted zeros and the explicit $1$ bit from the significand; Compute the actual exponent $e = e_\text{max} - k$;
        \label{decomp:normalize}
    \item Merge $s$, $e$, and the corrected significand back to an IEEE double-precision value.
        \label{decomp:ieee}
\end{enumerate}
As not the complete block has to be decompressed to retrieve one value, random access is possible.
To decompress value $c$ at index $i$, the only overhead is that $e_{max}$ must also be retrieved from the main memory.
However, the most efficient access is to read the whole block and reuse the cached $e_\text{max}$ to decompress all values in the block.

\subsection{Implementation and synthetic performance}
\label{sub:implementation}
To ensure good performance on the Nvidia H100 GPU, we perform the following optimizations:
(1) We utilize intrinsic functions to convert between the IEEE format and the appropriate integer type so we can analyze it bit-by-bit.
    The intrinsic function to count the leading zeros of an integer: \textit{count\_zero} is also mandatory for good performance.
(2) We mandate a block size $BS = 32$ for Nvidia GPUs.
    This allows us to use warp-shuffles, the fastest communication between threads, to determine $e_\text{max}$ during compression.
    Additionally, it guarantees that $e_\text{max}$ is cached for all threads of the warp during decompression of the same block.
(3) Have separate compression and decompression routines for $l = 2^x$ and $l \neq 2^x$.
    For $l=2^x$, compression and decompression are much simpler because values do not interleave in memory, making reading and writing them significantly faster.
(4) Perform all index computations in 32-bit integer types.
    Originally, we used 64-bit, but those are noticeably slower than 32-bit on an H100.
(5) Store the exponent for the blocks and the compressed values in separate memory locations, which simplifies the index computations substantially.

We implement the \frsz format in C++ and CUDA.
Additionally, for the decompression, we can utilize the Accessor interface in Ginkgo~\cite{cbgmres,accessorBLAS}, which is a software interface that decouples the storage format from the arithmetic format.
So far, it has been used to store values in half- or single-precision while performing all computations in double-precision.
The same interface is used for reading and decompressing data in \frsz while computing in double-precision because decompression does not require any form of communication with other threads.
We need to read the complete block of values to maximize cache usage, but that is already the access pattern for the Krylov vectors, so we do not need to treat them differently when we use \frsz.
Writing and compressing data can not be handled through the Accessor interface because it was designed for random access in both directions.
Our compression must be applied to a full block of values and requires local communication to find $e_\text{max}$.
Figure~\ref{fig:gmres} is the implemented GMRES algorithm, which also highlights all sections that compress or decompress data.

\begin{figure}[htb]
    \centering
    \includegraphics[width=\columnwidth]{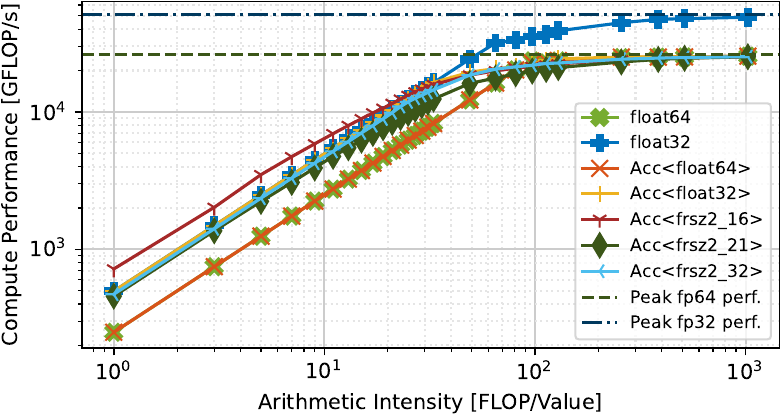}
    \caption{Performance on the H100}
    \label{fig:h100_performance}
\end{figure}

We evaluate the usage and efficiency of the \frsz decompression inside the Accessor interface with a synthetic benchmark that reads consecutive elements from the main memory and executes a pre-defined number of arithmetic operations on each value retrieved from main memory to vary the arithmetic intensity.
For each storage format, we run this benchmark for 27 arithmetic intensity settings.
In Figure~\ref{fig:h100_performance}, we report for increasing arithmetic intensity the performance of the kernel using different storage formats.
We use an array with $2^{28}$ randomized elements to utilize the full GPU.
Each data point in the plot is the minimum from 10 individual executions.
The resulting roofline analysis allows us to compare the performance and memory bandwidth of the different storage and compression formats.
We note that \double and \float do not use the Accessor interface but read and compute in their respective precision directly.
This allows us to evaluate the overhead of the Accessor.
\textit{\Acc{\doubleN}} and \textit{\Acc{\floatN}} compute in double-precision while storing the values in \double and \float, respectively.
The performance of the Accessor is identical to the native implementation as long as they are memory-bound, which proves the zero-cost abstraction.
For \frsz, we always use $\text{BS} = 32$ and three different bit lengths: $l\in\{16,21,32\}$.
We chose $16$ and $32$ for their value alignment, making their decompression less complex, and $21$ to observe the penalty of the additional decompression steps.


$l=16$ shows the highest performance per value. However, it is not a factor of $2$ faster than the single-precision storage, which means we do not saturate the full bandwidth.
Additionally, the gap between \double and \textit{\Acc{\frszaccN{16}}} decreases rapidly with higher arithmetic intensity, so it is unsuitable for workloads with slightly higher arithmetic intensity.
\textit{\Acc{\frszaccN{32}}} achieves slightly lower performance than \textit{\Acc{\floatN}}.
The reason is simple: \frszacc{32} needs 33 bits per value on average because it needs to store one exponent value, which occupies 32 bits as well, per block, and since $\text{BS} = 32$, the average bits per value is: $(\text{BS} \cdot l + 32) / \text{BS} = (32 \cdot 32 + 32) / 32 = 33$.
This difference could be explained by the additional exponent that needs to be read by block, which translates to roughly $33$ bits per value for \frszacc{32}.
When measuring the bandwidth instead of performance, \textit{\Acc{\frszaccN{32}}} reaches $1991 \text{GB/s}$, which is $\approx99.6\%$ of the reachable bandwidth.
This confirms the viability of our compression target: Our decompression algorithm can saturate the bandwidth and has cycles to spare for many additional floating-point computations.

\textit{\Acc{\frszaccN{21}}} displays a similar performance to \textit{\Acc{\frszaccN{32}}} despite reducing the memory footprint by $\approx33\%$.
This clearly states that the overhead in the more complex index computation and the unaligned memory read operation is too high to translate to higher performance.
\frszacc{21} will always be less precise than \frszacc{32}, which means \frszacc{21} is only useful in case \frszacc{32} would not fit in GPU memory.
We have not encountered any problem large enough to reach the GPU memory capacity, so its usefulness is limited.

%% file: sections/methodology.tex
\section{Methodology}
\label{sec:methodology}

This section describes common details to reproduce our experiments and motivations for these choices. We begin with hardware and software choices, then discuss the choice of benchmark problems for evaluation, and finally, we discuss the configuration of compressors used in the comparison.

\subsection{Hardware and Software}

We choose the latest NVIDIA H100 GPU for our experimental evaluation. 
We also considered other GPUs, but we present results only for the H100 for brevity. The results for other GPUs are not meaningfully different in conclusion.
The Nvidia H100 is the PCIe variant with 80 GB of RAM, 50 MB of L2 cache, 25.6 TFLOP/s double-precision, and 51.2 TFLOP/s single-precision performance.
The peak memory bandwidth is 2000 GB/s.
The host system is a server with two Intel Xeon Silver 4309 processors.

We use the default compilers on the system: CUDA 12.1 and GCC 11.4.
We choose the default (latest) packages from spack.
We utilize LibPressio~\cite{underwood_productive_2021} version 0.98.0 to manage and interact with the other compression algorithms we use: sz version 2.1.12.5, sz3 version 3.1.7 and zfp version 1.0.0.

Our CB-GMRES implementation with \frsz and the benchmark code is open-source and can be accessed in the Ginkgo branch \texttt{2024-drbsd-paper}\footnote{\url{https://github.com/ginkgo-project/ginkgo/tree/2024-drbsd-paper}}.

\subsection{Problem Selection}

\begin{table}
\centering
\begin{tabular}{lrrr}
\hline
\textbf{Matrix} & \textbf{Size} & \textbf{Non-zeros} & \textbf{target RRN} \\
\hline
atmosmodd & 1,270,432 & 8,814,880 & $4.0\cdot10^{-16}$ \\
atmosmodj & 1,270,432 & 8,814,880 & $4.0\cdot10^{-16}$ \\
atmosmodl & 1,489,752 & 10,319,760 & $4.0\cdot10^{-16}$ \\
atmosmodm & 1,489,752 & 10,319,760 & $4.0\cdot10^{-16}$ \\
cfd2 & 123,440 & 3,085,406 & $1.8\cdot10^{-10}$ \\
HV15R & 2,017,169 & 283,073,458 & $1.6\cdot10^{-02}$ \\
lung2 & 109,460 & 492,564 & $1.8\cdot10^{-08}$ \\
parabolic\_fem & 525,825 & 3,674,625 & $4.0\cdot10^{-16}$ \\
PR02R & 161,070 & 8,185,136 & $4.0\cdot10^{-03}$ \\
RM07R & 381,689 & 37,464,962 & $8.0\cdot10^{-03}$ \\
StocF-1465 & 1,465,137 & 21,005,389 & $4.0\cdot10^{-06}$ \\
\hline
\end{tabular}
\caption{Details of the computational fluid dynamic matrices used from SuiteSparse}
\label{tab:matrices}
\vspace{-.5cm}
\end{table}

The matrices we use are from the SuiteSparse matrix collection~\cite{suite_sparse}.
It is widely used for various sparse benchmarks because of its vast size (2,893 matrices) and diversity in domains (48).
We focus on matrices that solve computational fluid dynamics problems because they worked poorly in~\cite{cbgmres} and aim to improve the convergence rate with our \frsz compression.
As an additional restriction, they need to have a matrix size of more than $100,000$ rows to avoid caching mechanisms blurring the understanding of the performance analysis.
The most important properties of the matrices we use can be seen in table~\ref{tab:matrices}.

For each matrix $A$, we generate the right hand side $b$ deterministically and identical to \cite{cbgmres} to ensure fair comparisons:
First, we take a vector $s$ and set the $i$-th entry to its sin value: $s[i] = \text{sin}(i) \: \text{for}\: i\in \{{0,1,\dots,n-1}\}$.
The expected solution $x_\text{sol}$ is gained by normalizing $s$ with its unit norm: $x_\text{sol} = s / \|s\|_2$.
The right-hand side $b$ is computed by multiplying the expected result with the matrix A: $b = A\cdot x$.
All GMRES algorithms are started with the initial guess $x_0 = \vec{0}$, using a restart parameter $m=100$\footnote{To limit the memory requirements, GMRES is restarted after a Krylov basis of 100 vectors has been built up. The restart uses the latest solution approximation as initial guess, and starts building up a new Krylov search space.}, and are stopped when the solution approximation $x$ fulfills the relative residual norm specified in Table~\ref{tab:matrices}: $\|A\cdot x-b\|_2 \le \text{RRN}\cdot\|b\|_2$.

\subsection{Solver Configuration}
GMRES is an iterative solver that tries to solve the equation $A\cdot x = b$ to a predefined accuracy.
The sparse matrix $A$ and the right-hand-side vector $b$ are problem-dependent and immutable inputs to the solver.
GMRES also needs an initial guess $x_0$ as a starting vector, which is then iteratively improved to get closer and closer to the solution $x_\text{sol}$.
The residual $r$ is the difference vector between the targeted result $b$ and the result achieved with the current approximation of $x$.
To quantify the quality of the current approximation of $x$ in a single number, we compute the relative residual norm (RRN):
\begin{equation}
\label{eq:rel_res_norm}
\text{RRN} = \frac{\|r\|_2}{\|b\|_2} = \frac{\|b - A\cdot x\|_2}{\|b\|_2} 
\end{equation}
This value is given to the GMRES algorithm to determine when the computed solution approximation is sufficiently accurate.
The lower RRN, the closer we are to the exact solution.
The ideal case is $\text{RRN} = 0$. However, we compute with finite IEEE double-precision, with a unit roundoff of $u\approx10^{-16}$, thus $\text{RRN} = 0$ may not be achievable. 
Additionally, some problems are inherently difficult to solve, so we adjust our target accuracy for each problem.
For this, we solve each problem with $20,000$ iterations of a standard double-precision GMRES.
The solution accuracy achieved is then used with some wiggle room as the stopping criterion for the CB-GMRES variants using different storage formats for the Krylov basis. 

Table~\ref{tab:matrices} lists the obtained targeted relative residual norms we use from this point forward.

We do not use any preconditioner to not blur the numerical impact by the use of a sophisticated preconditioner.


\subsection{Compression Configuration}

Details about the \frsz compression are outlined in Section~\ref{sec:design}.
Every decompression happens through the Accessor interface, while the compression is called directly without an intermediate interface.
\textit{\frszacc{XX}} corresponds to the \frsz format with $\text{BS}=32$ and $l = XX$.
We also experimented with different block sizes, but the end-to-end runtime worsens with block sizes different than 32 elements, so we focus purely on $\text{BS} = 32$.

We also run all experiments with the original CB-GMRES storage formats:
\double, which stores the values in IEEE double-precision storage format;
\float, which stores values in single-precision;
and \half, which store values in half-precision.
All these options use the Accessor to have varying storage formats, but all arithmetic calculations are still done in IEEE 754 double precision.

While we also want to evaluate the compression efficiency of other compression schemes, implementing these in the Accessor interface would require a substantial amount of work. Thus, we decided to simulate the effect of other compression schemes on the CB-GMRES convergence by using them via LibPressio~\cite{underwood_productive_2021}. 
We do this by compressing and immediately decompressing the Krylov vectors through the LibPressio interface.
This helps us to analyze the loss of information of various compressors without the need to implement any of them.
We focus on SZ, SZ3, and ZFP because they are leading lossy compressors with multiple error bounds supported while using different decorrelation strategies.

We experiment with many error-bound settings for SZ, SZ3, and ZFP.
Many behave the same or at least very close to the others, so we chose the settings shown in Table~\ref{tab:compressor_settings}.

\begin{table}[h]
\centering
    \begin{tabular}{l c r}
        \hline
        \textbf{Name} &  \textbf{error-bound type} & \textbf{error-bound} \\
        \hline
        sz3\_06 & absolute & $10^{-06}$ \\
        sz3\_07 & absolute & $10^{-07}$ \\
        sz3\_08 & absolute & $10^{-08}$ \\
        \hline
        zfp\_06 & absolute & $1.4\cdot 10^{-06}$ \\
        zfp\_10 & absolute & $4.0\cdot 10^{-10}$ \\
        \hline
        sz\_pwrel\_04 & relative & $10^{-04}$ \\
        sz3\_pwrel\_04 & relative & $10^{-04}$ \\
        \hline
        zfp\_fr\_16 & fixed rate & 16 bits \\
        zfp\_fr\_32 & fixed rate & 32 bits \\
        \end{tabular}
    \caption{Compressor name and requested bounds.}
    \label{tab:compressor_settings}
    \vspace{-.5cm}
\end{table}

%% file: sections/evaluation.tex
\section{Evaluation} \label{sec:evaluation}

There are two classes of experiments to perform: comparisons of the quality of the solution and end-to-end performance. 
GMRES requires a compression format that reduces data transfers without impacting neither the accuracy of the final result nor requiring significantly more iterations to reach convergence.
Lastly, we consider the total algorithm runtime.

\subsection{Quality of Solution}
\label{sub:eval_accuracy}

\begin{figure}[htb]
    \centering
    \includegraphics[width=\columnwidth]{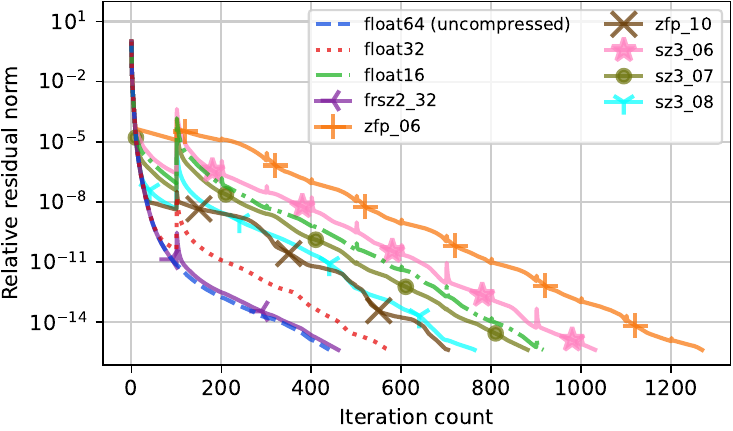}
    \caption{Residual norm development for the \textit{atmosmodd} matrix with various compressions.}
    \label{fig:residual_history_atmosmodd_other}
\end{figure}

First, we compare the accuracy of \frsz with the other compression schemes presented in~\cite{liang_sz3_2023,lindstrom_fixed-rate_2014,liang_efficient_2018} for different absolute error-bound settings to understand their impact on conversion rates.
We choose the matrix \textit{atmosmodd} here because this is one of the rare test problems where storing the Krylov basis in single precision impacts the accuracy of the final result~\cite{cbgmres}.
Figure~\ref{fig:residual_history_atmosmodd_other} displays the relative residual norm development throughout the GMRES solve.
For \textit{atmosmodd}, we target a relative residual norm of $4\cdot10^{-16}$.
This target is reached faster if the compression preserves more information.

For \textit{atmosmodd}, the convergence rates for the different compression schemes vary substantially. 
The convergence rate of \frszacc{32} is close to matching the convergence rate of the uncompressed \double storage.
\frszacc{32} improves over \float despite using the same space.
We can attribute this improvement in convergence over \float to the increased space to store precision information created by externalizing the exponent to the block.

Considering the other absolute error bounded compressors, none of the ZFP or SZ3 settings manage to match the convergence of the \float compression, even though \textit{sz3\_08} uses 46 bits per value on average, compared to the 32 from \float.
\textit{zfp\_10} outperforms \textit{sz\_08} both in convergence and compression rate as it only uses 28 bits per value.
We attribute the slower convergence of SZ and ZFP to these compressors' ill-fated attempts to predict or decorrelate the uncorrelated Krylov vectors, resulting in a bias in the reconstructed values.

\begin{figure}[htb]
    \centering
    \includegraphics[width=\columnwidth]{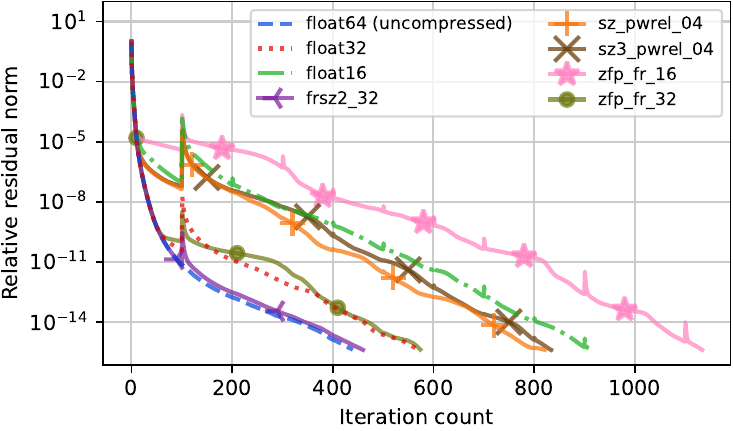}
    \caption{Residual norm development for the \textit{atmosmodd} matrix with pointwise relative error settings.}
    \label{fig:residual_history_atmosmodd_pwrel}
\end{figure}

Next, we investigate the pointwise relative error bounds in Figure~\ref{fig:residual_history_atmosmodd_pwrel}. We observe that the pointwise error bounds enable better convergence rates than absolute error bounds.
The pointwise relative error preserves $x(1 - \epsilon) \leq \Tilde{x} \leq x(1 + \epsilon)$.
Consequently, the values' magnitude is better preserved than using the absolute error bound, which is more similar to our \frsz approach.
Still, though the convergence rates are improved, none of the compressors can match the\float compression in terms of GMRES convergence.
The fixed-rate ZFP compression achieved the best convergence rate among the other compressors.
It occupies the same memory as \float but retains slightly less information for this application.
\frszacc{32} has the best convergence rate among all tested compression techniques.

\begin{figure}[htb]
    \centering
    \includegraphics[width=\columnwidth]{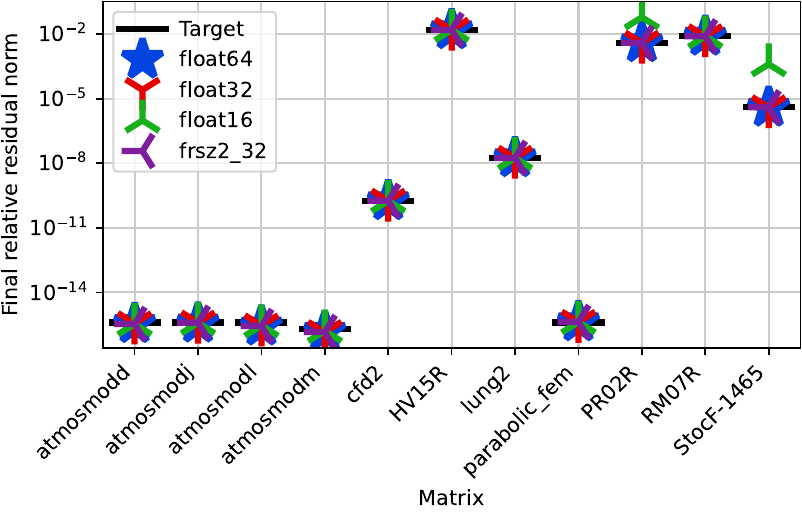}
    \caption{Final relative residual norm for various matrices on the H100.}
    \label{fig:rrn_h100}
    \vspace{-.5cm}
\end{figure}

After showing that \frszacc{32} improves convergence for one test problem, we investigate whether this effectiveness generalizes to other problems.
Figure~\ref{fig:rrn_h100} presents the target and the achieved relative residual norm for the storage formats \double, \float, \half, and \frszacc{32} for all considered matrices.
The exact target relative residual norm is listed in Table~\ref{tab:matrices}.
We observe two instances where we do not reach the targeted relative residual norm with \half: \textit{PR02R} and \textit{StocF-1465}.
The loss of information is too significant for these problems.
For the other matrices, all settings converge to the target precision.

\begin{figure}[htb]
    \centering
    \includegraphics[width=\columnwidth]{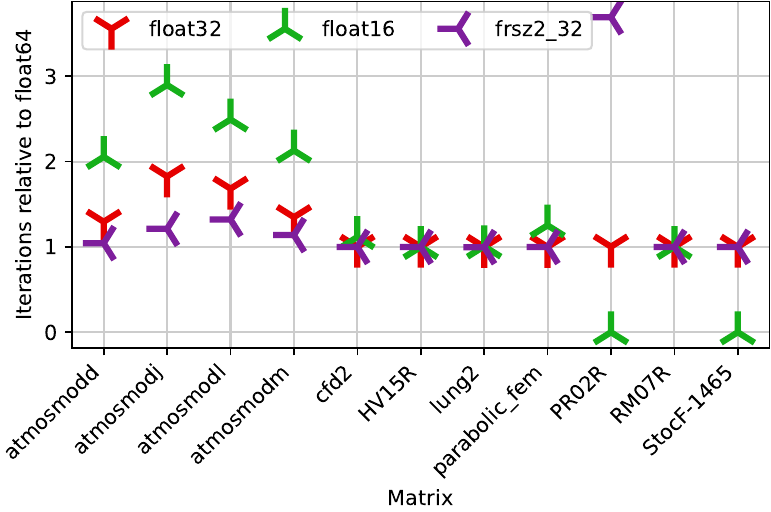}
    \caption{Mean number of iterations to the solution of various matrices for the H100 over 10 runs. Zero means the solver does not reach the target precision.}
    \label{fig:iterations_h100}
\end{figure}

Next, we compare the convergence rate with the \double convergence rate.
We do this by showing the number of iterations each storage type needs to achieve the target precision as a factor of the reference \double in Figure~\ref{fig:iterations_h100}.
We set the relative value to zero if the target accuracy is not achieved.
We observe that all matrices with the prefix \textit{atmosmod} behave similarly: \double converges fastest, followed by \frszacc{32}, then \float, and finally \half.
Here, \frszacc{32} is clearly the best compression format because it comes with the smallest iteration overhead.
In contrast, \textit{PR02R} is the worst problem for \frsz.
\frszacc{32} eventually converges to the target norm, but the iteration count increases by $3.5\times$.
All other matrices barely show a difference in convergence rate.

\begin{figure}[htb]
    \centering
    \begin{subfigure}[b]{\columnwidth}
        \includegraphics[width=\columnwidth]{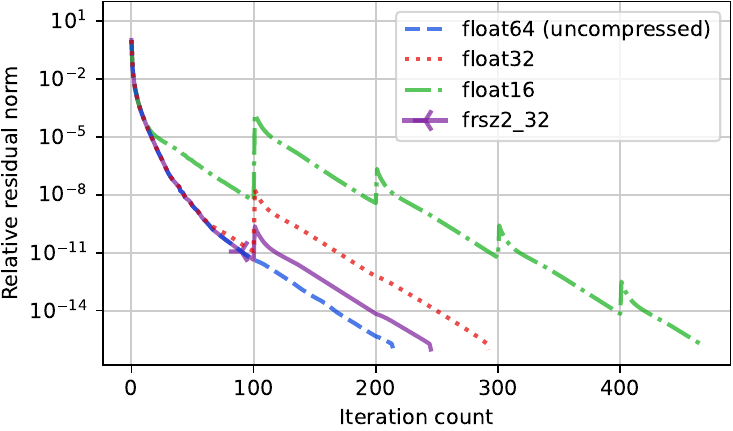}
        \caption{Matrix \textit{atmosmodm}}
        \label{fig:residual_history_atmosmodm}
    \end{subfigure}%
    \\
    \begin{subfigure}[b]{\columnwidth}
        \includegraphics[width=\columnwidth]{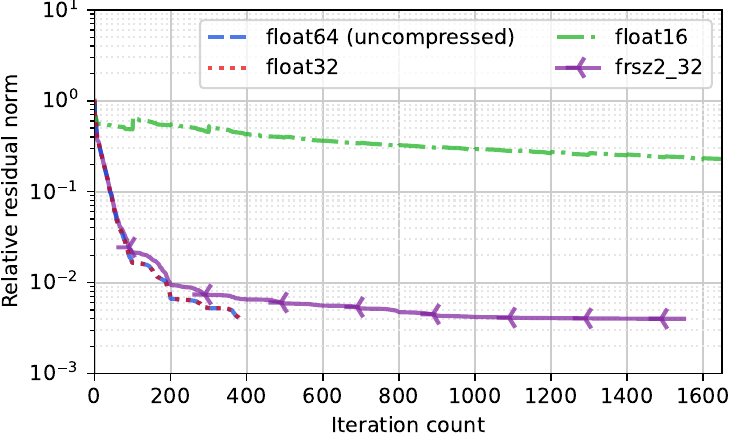}
        \caption{Matrix \textit{PR02R}}
        \label{fig:residual_history_PR02R}
    \end{subfigure}%
    \caption{Relative residual norm development for the best- and worst-performing matrices for \frsz.}
    \label{fig:residual_history}
    \end{figure}

We now focus on the test cases where \frszacc{32} performs extremely well and extremely badly. 
Figure~\ref{fig:residual_history} provides a deeper insight into the convergence rate by plotting the relative residual norm for each iteration for the matrices where \frszacc{32} works well, represented by \textit{atmosmodm} in Figure~\ref{fig:residual_history_atmosmodm}, and the matrix \textit{PR02R} in Figure~\ref{fig:residual_history_PR02R}.

\textit{atmosmodm} has a big residual norm correction after the first restart at iteration $100$ for all storage formats except for the uncompressed \double.
These corrections exist because the residual norms are only explicitly computed at every restart in GMRES, which we do every $100$ iterations.
For all other iterations, it only adjusts the previous residual norm by the assumed amount of improvement.
During the restart, the residual and its norm are explicitly computed and used as the new baseline.
These adjustments are the sudden jumps in Figure~\ref{fig:residual_history}.
\frszacc{32} seems to recover from that correction the fastest from all the other compressions and only requires $31$ more iterations to achieve the same accuracy as the uncompressed at convergence.
The order from best to worst seems sorted by the number of significand bits for each compression scheme.

Figure~\ref{fig:residual_history_PR02R} shows a different side of the compression scheme.
Here, \frszacc{32} follows both the single- and double-precision storage format until a relative residual norm of $2\cdot 10^{-2}$ is reached, then stagnates.
It barely improves the residual norm between iterations $400$ and $1600$, which might indicate that it reached its maximum accuracy.
Half-precision does not even reach an accuracy of $10^{-2}$.
Even after $20,000$ iterations, it only managed to go down to $5\cdot 10^{-1}$.

\begin{figure}[htb]
    \centering
    \includegraphics[width=\columnwidth]{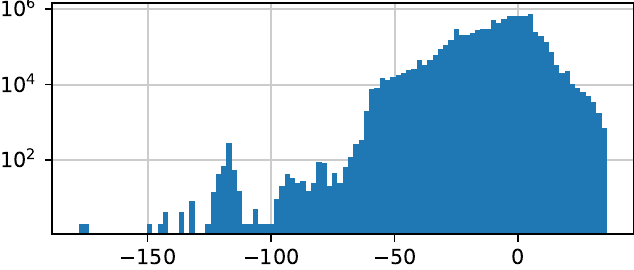}
    \caption{Base-2 exponent histogram of all non-zero values of \textit{PR02R}.}
    \label{fig:exp_hist_PR02R}
    \vspace{-.5cm}
\end{figure}

Part of the reason might be the huge range of non-zero values the matrix \textit{PR02R} has.
Figure~\ref{fig:exp_hist_PR02R} visualizes the exponent distribution of matrix \textit{PR02R}, which ranges from $-178$ to $36$.
If a block of Krylov basis values contains exponents with a large range, we lose a lot of precision in values with smaller exponents when we fill the significands with zeros in the normalization step.
However, this is not the only contributing factor.
The matrix \textit{HV15R} has an extremely similar value distribution to \textit{PR02R}.
The ordering of non-zero values in \textit{HV15R} may lead neighboring Krylov vector values to have a similar magnitude, mitigating the effects observed in \textit{PR02R}.

\subsection{End-to-End Performance}
\label{sub:eval_performance}

\begin{figure}[htb]
    \centering
    \includegraphics[width=\columnwidth]{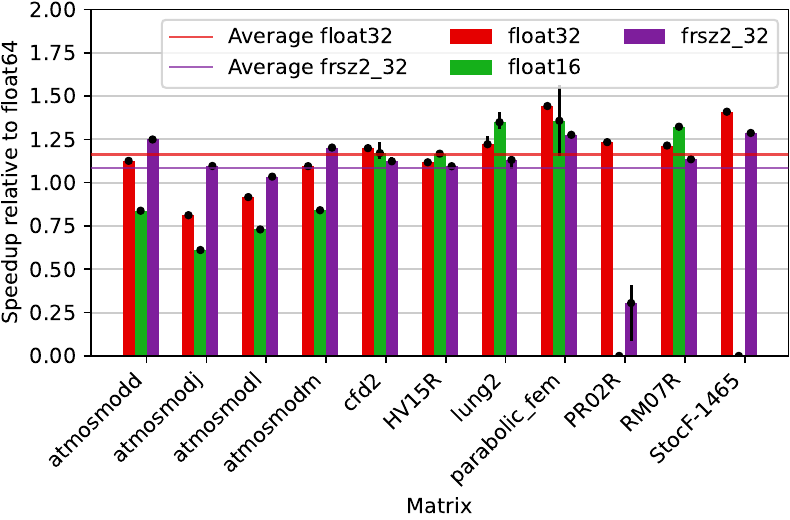}
    \caption{Mean speedup of various matrices for the H100 with an error bar. Each matrix was solved ten times.}
    \label{fig:speedup_h100}
\end{figure}

Finally, we investigate the end-to-end speedup achieved when using \frszacc{32}.
Looking back at the read performance in Figure~\ref{fig:h100_performance} and combining that with the usually same number of iterations in Figure~\ref{fig:iterations_h100}, we expect the \frszacc{32} performance to be similar to using single precision for compression.
The mean speedup compared to double-precision storage is shown in Figure~\ref{fig:speedup_h100} with error bars.
The entire bar is removed from a matrix if a storage format does not reach the targeted relative residual norm.
As expected, \frszacc{32} performs well in the \textit{atmosmod} group.
It is faster than single-precision storage and also beats the double-precision storage format.
For problems outside the \textit{atmosmod} group, \frszacc{32} is consistently slower than the \float storage format.

The average speedup over all matrices for the \float storage format is $1.16$, while the average is $1.09$ for \frszacc{32}.
Ignoring matrix \textit{PR02R}, the average speedup increases to $1.16$.

We also experimented with a variant of our method using 21 bits, the \frszacc{21} storage format.
Experiments revealed that the convergence for \frszacc{21} is superior to \half, but dramatically slower than \frszacc{32} due to worse alignment.
We, therefore, omit the experimental results from the evaluation.

%% file: sections/related_work.tex
\section{Related Work}\label{sec:related_work}


There is a rich history of using lossy compression to accelerate computations.
One paper used lossy compression to effectively expand the memory by compressing key data structures instead of recomputing them when they could not all fit in memory \cite{gok_pastri_2018}.  
A more recent paper used compression to speed up I/O \cite{wu_full-state_2019} even if the resulting operations were slower to make it possible on a given set of resources.
For accelerating memory-bound algorithms, the compression has to happen in processor registers, without touching main memory for either compression or decompression. In both of these papers, the operation that was avoided by using compression was substantially slower than a single access from the HBM on the GPU.  We in contrast present a uniquely challenging case where the operation being modified is itself fast.

The idea of in-register data compression to speed up memory-bound linear algebra operations was initially realized by casting data to lower precision. In \cite{adaptivejacobi}, the authors accelerate a block-Jacobi preconditioner by storing the individual block-inverses in lower precision, while keeping high precision for the arithmetic operations. The same strategy was later used for accelerating sparse approximate inverse preconditioners~\cite{compressedisai}. Similarly, 
the idea of compressing the Krylov basis of a GMRES iterative solver was initially using low precision for compression of the vector values~\cite{cbgmres}. Almost at the same time, this concept was proposed also by \cite{agullocbgmres}, however following a more sophisticated strategy by storing the preconditioned Krylov vectors inside a flexible GMRES solver in low precision. This improves the numerical stability at the price of reduced runtime benefits.
Casting to lower precision can render only a moderate compression ratio without losing too much information. This motivates the use of sophisticated compression strategies. In \cite{jlescreport} we investigated the use of advanced compression strategies yielding large compression factors. The results, however, indicated that these methods are too clumsy to operate on GPU registers, and the compression factors were too small to translate to speedup factors. We hence developed a more lightweight compression drawing a good balance between compression ratio and compression cost and presented the results in this paper.

However, the alternative was not to perform the computation at all.  In GMRES, the inputs are 1) not bound by GPU memory size indicating the problem would be unlikely if it exists, and 2) have the clear alternative of performing the calculation in mixed or lower precision such as Float32 achieving a similar outcome without the possible overheads of compression.


%% file: sections/conclusion.tex
\section{Conclusions and Future Work}\label{sec:conclusions}

In this paper, we present \frsz, a highly specialized compressor for GMRES, that provides unparalleled performance among modern compressors and is uniquely capable of accelerating end-to-end performance of GMRES for a class of applications up to $1.3\times$ compared to uncompressed methods as well as $1.2\sim3.1\times$ faster than existing compressors obtaining 99.6\% of the peak bandwidth at the roofline.

However, more work is needed to realize \frsz as a generalizable solution for use in GMRES solvers in packages such as Ginkgo.  Either 1) there needs to be continued work to accelerate \frsz even further -- this could come in the form of additional hardware improvements for certain assembly instructions (i.e., masked shuffle operations) used in decompression routines or algorithmic improvements that could eliminate the dependence on these slow instructions or changes to the balance between memory and compute bandwidth 2) we need an accurate, robust, and fast method to predict when an application will benefit from \frsz compared to mixed-precision methods.

Given the degree of optimization already applied to \frsz, we believe these benefits are most likely to come from predictions that can be applied just before the first restart.  We explored this briefly in our work prior to submission. We considered features such as the condition number, value distribution, exponent distribution, and even autotuned methods that detect and observe the convergence per unit time of several candidate methods and then speculatively execute that the best initial method will continue to dominate.  We have only scratched the surface of possible methods, and with further work, an appropriate prediction method may be identified.

